\documentclass[10pt,aps,pra,showpacs,amssym,amsfonts,amsmath,floatfix,]{revtex4}
\usepackage{epsfig}
\usepackage{bm}
\usepackage{graphicx}
\usepackage{float}
\usepackage{latexsym}

\begin{document}
\title{High-order harmonic generation from C$_{20}$ isomers}
\author{F. Cajiao - V\'elez, A. Jaron}

\begin{abstract}
High-order harmonic responses from three $C_{20}$ isomers: fullerene, ring, and bowl, are calculated
within the modified Lewenstein model for molecular systems. Spectra for all three structures exhibit
intense modulations of the harmonic spectrum along the plateau and some of them can be interpreted
as a consequence of multi-center interference effects. Each structure shows characteristic modulation
patterns in peak harmonic intensities, which are directly related to zeroes in the recombination
matrix element as a function of the three components of momentum. Different C20 isomers lead
to different harmonic polarizations depending on the geometric configuration of carbon atoms and
molecular orientation. 

\end{abstract}

\maketitle

\section{Introduction}

In recent years, high-order harmonic generation (HHG) has been a very active research topic in experimental and theoretical fields due to the new and interesting properties of the non-linear interaction between laser radiation and molecules or atoms. This process involves the transformation of multiple low-energy photons coming from the laser field into a single high frequency photon \cite{Ciappina_JB}. When the intense laser pulse interacts with a
molecule or atom, coherent radiation of frequencies that are integer multiples of the original driving one are emitted \cite{Lewenstein1}. HHG spectra exhibit particular characteristics as a consequence of the non-linear interaction. The first harmonics decrease in intensity rapidly before a plateau, where the strength of the peaks is fairly constant. A sharp cutoff determines a new region characterized by a fast decrease of the harmonic signal. The maximum energy at the end of the plateau can be approximated by the formula~\cite{Corkum,Krause,Lewenstein1}
\begin{equation}
E_{\rm cutoff}=I_p+3.17 U_p \,, \label{eq:cutoff}
\end{equation}
where $I_p$ is the ionization potential and $U_p=e^2\mathcal{E}_0^2/(4\omega_0^2)$ corresponds to the ponderomotive energy of a free electron driven by a monochromatic plane wave. Here, $\mathcal{E}_0$ represents the amplitude of the electric field and $\omega_0$ is the laser carrier frequency. Atomic units are used along this paper while the electron charge ($e<0$) is explicitly written in the formulas. For numerical calculations we set $|e|=1$, unless stated otherwise.

As the HHG process is of purely quantum mechanical origin, the most
suitable treatment comes from the complete solution of the time-dependent Schr\"odinger equation (TDSE) which is a prohibitive task for multielectron systems interacting with intense laser fields.  Approximate methods such as time-dependent density functional theory (TDDFT) (see, e.g.,~\cite{Ishikawa1,Ishikawa2,Lappas,Kamta,Yuqing}) and time dependent Hartree-Fock theory (see e.g., ~\cite{HF}) have been applied to intermediate size systems. 
Note that to solve the TDSE or the TDDFT/TDHF is a challenging task already for two-electron systems. While it is time consuming and requires a big amount of computational effort, those methods can be used to model the interaction of atoms and simple molecules with strong laser fields. Their solution presents, with good agreement with experiments, a plateau before the sharp cutoff~\cite{Kulander}. With increasing the laser field intensity or the complexity of the molecule, solving the TDSE or the TDDFT/TDHF becomes prohibitive~\cite{Chirila,Lein2}.

Many features of the HHG process can be understood by means of the semiclassical three-step model. In the first step, the electron is ionized by tunneling effects due to the distortion of the atomic potential by the oscillating electric field. In the second
step, the electron travels in the continuum as a classical particle subject to the Lorentz forces in the laser field. Finally, the electron recombines with the parent ion with the consequent emission of a highly energetic photon~\cite{Corkum}. Due to the classical treatment of the propagation step, many quantum-mechanical effects, including the electron wave packet spreading in the continuum and its acquired phase, are ignored. 

The Lewenstein model offers a complete quantum-mechanical treatment of HHG by considering the electron wave function, between ionization and recombination events, as a dressed plane wave propagating in the laser field. The Coulomb interaction between parent ion and ejected electron is neglected, so the total evolution in the continuum is governed by a Volkov type evolution operator.

When applied to atoms, the Lewenstein model has proven to give a very good quantitative and qualitative agreement with ab-initio calculations~\cite{Chirila} with the advantages of being
considerably faster and requiring less computational effort. Such model has been extended and applied to polar and non-polar molecules
\cite{Chirila,Etches,Etches2,Kanai}.

An important feature of the Lewenstein model is its absence of gauge
invariance and many problems related to the gauge choice may arise, specially when the HHG spectrum from molecules is calculated. For example, if the harmonic emission from an atom displaced from the origin of coordinates is considered, new effects are observed in the response calculated in the length and velocity gauges. Both of
them show the presence of even and odd peaks, but in the length gauge the odd harmonics are not invariant under spacial translation~\cite{Chirila}. All those drawbacks seem to be partially avoided by the insertion of new terms into the semiclassical action (see, Sec. \ref{sec:theory}). Those terms account for the multi-center features of a molecule and predict
interference modulations along the plateau~\cite{Chirila}. In Ref.~\cite{Chirila} it has been shown that the notorious effects related to the length gauge choice (e.g., nonphysical enhancement of the plateau and increasing harmonic intensities with the internuclear distances) are less important for small molecular sizes. If the nuclear separation is shorter than the quiver radius of the electron [$\alpha_0=|e|\mathcal{E}_0/\omega_0^2$], the artifact features seem to be less intense. This has been proven for diatomic molecules and extended to bigger systems~\cite{Ciappina_JB,Chirila}. When a large number of
atoms is taking into consideration, it is useful to define a parameter $Q$,
\begin{equation}
Q=\frac{R_{\rm max}}{2\alpha_0}\,. \label{eq:q}
\end{equation}
Here, $R_{\rm max}$ is the maximum distance between two
nuclei in the molecule and $\alpha_0$ (the quiver radius) is the amplitude of classical oscillations of an electron in the laser field. This parameter may help to determine whether or
not the length gauge can be applied without introducing nonphysical effects (see, Ref.~\cite{Ciappina_JB}).

The Lewenstein model predicts important contributions to the harmonic response from two different electron trajectories (\textit{long trajectory} and \textit{short  trajectory}~\cite{Carlos}). Such paths are determined by the time spent by the electron in the continuum. It has been proven that different  trajectories can interfere constructively or destructively, generating strong modulations on the harmonic peak intensities along the plateau~\cite{sansone,gkortsas}.

\begin{figure}
\includegraphics[width=\columnwidth]{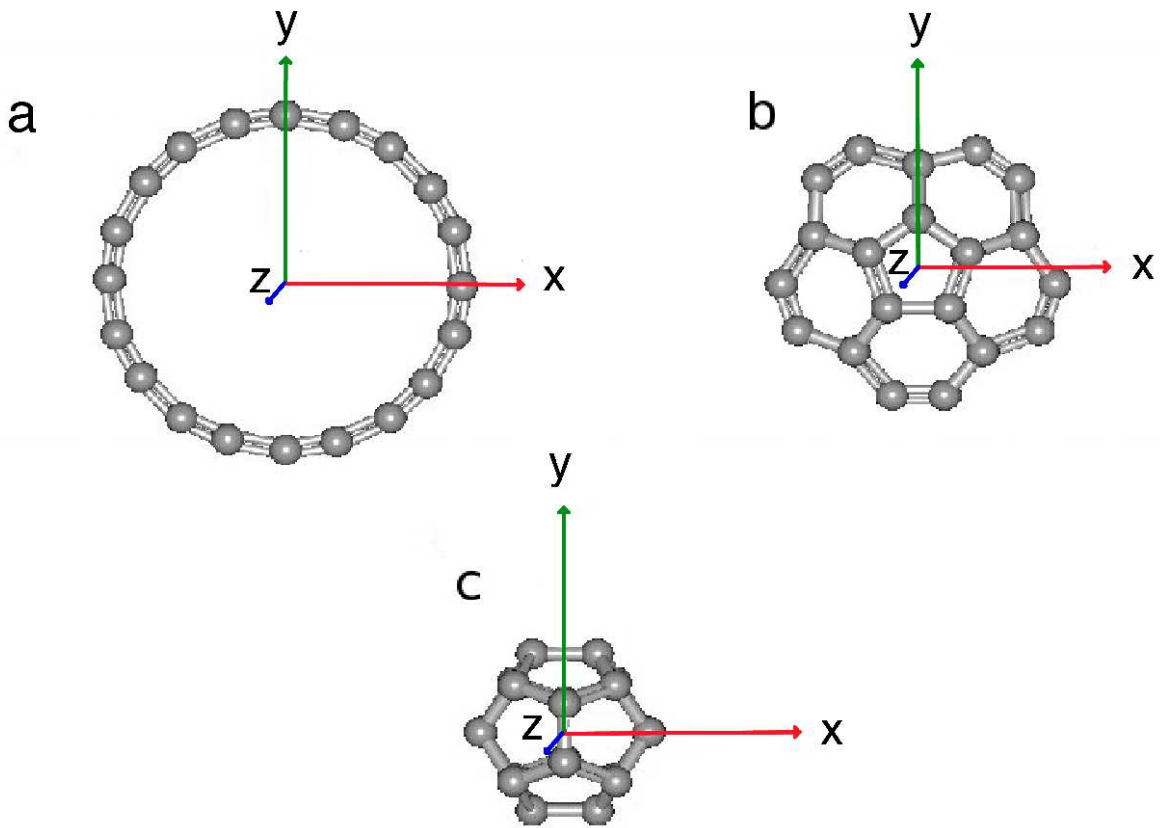}
\caption{Structure of three $C_{20}$ isomers. The carbon atoms distributions were obtained from the standard computational chemistry package GAMESS using the 6-311G basis set. The structures correspond  to (a) the monocyclic ring, (b) the bowl, and (c) the fullerene (also known as the cage). }
\label{fig:all}
\end{figure}

The present paper is devoted to the analysis of the harmonic
response from three different $C_{20}$ isomers: the fullerene (or cage), the ring, and the bowl (see, Fig.~\ref{fig:all}). The cage is the smallest possible fullerene, with atomic centers located at the corners of twelve pentagons. The bowl geometry can be considered as a fragment of the \textit{buckminsterfullerene} $C_{60}$. The planar monocyclic ring is another configuration adopted by the $C_{20}$ family. The symmetry and planar
structure of such molecule, together with its particular electronic
configuration, make the $C_{20}$-ring an interesting harmonic target. Due to the highly symmetric nature of the fullerene and ring, no permanent dipole moment is observed in such structures. As a consequence of the lack of an inversion center, the bowl is characterized by a permanent electric dipole moment pointing along the $z$-direction. In Table~\ref{tab:ionization_potentials}, we present the values of the  mentioned static dipole moments together with the calculated ionization potentials of the Highest Occupied Molecular Orbital (HOMO), ($Ip_{\rm H}$) and HOMO-1 ($Ip_{{\rm H}-1}$). In order to calculate the molecular orbitals for each isomer, Hartree-Fock methods included in the standard quantum chemistry package GAMESS \cite {gordon} are used after a proper convergence is guaranteed. In Pople's notation, the basis set used in the calculations is written as 6-311G (see, e.g.,~\cite{bib:pople,bib:pople2}).

\begin{table}
\begin{tabular}{|c|c|c|c|c|c| }
\hline
{Isomer} &$Ip_{\rm H-1}$ (a.u.) & $Ip_{\rm H}$ (a.u.) & $d_x$ (D) & $d_y$ (D) & $d_z$ (D) \\\hline
{ Ring } &$0.3209$ & $0.3192$ & $0.0$ & $0.0$ & $0.0$ \\\hline
{ Bowl } &$0.3658$ * & $0.3568$ * & $0.0$ & $0.0$ & $0.2048$ \\\hline
{ Fullerene } &$0.3293$ & $0.2782$ & $0.0$ & $0.0$ & $0.0$ \\\hline
\end{tabular}
\caption{Ionization potentials and static dipole moments for three stable $C_{20}$ structures. The data was obtained by Hartree-Fock methods according to the standard 
quantum chemistry package GAMESS with the 6-311G basis set. The asterisk (*) denotes a two-fold degeneracy of the molecular orbital. In this Table we only present
ionization potentials of HOMO and HOMO-1, which are the most important for the present analysis.}
\end{table}\label{tab:ionization_potentials}

 In order to illustrate the effects of multi-center and quantum path interferences, a relatively large intensity of $I=5\times 10^{14}{\rm
 W}/{\rm cm}^2$ and a wavelength of $800{\rm nm}$ have been chosen for the calculations. This guarantees that the location of the cutoff is beyond the $60$th harmonic order for all structures. The same analysis can be performed for lower intensities and larger frequencies, so the plateau is long enough to observe the interference effects on the harmonic response. In Table \ref{tab:q} we show the average radius of each $C_{20}$ isomer together with the corresponding parameter $Q$. As it can be seen, the small values of $Q$
suggest that the length gauge formalism can be applied to analyze the harmonic spectrum from the three $C_{20}$ structures, given the aforementioned laser parameters, without introducing nonphysical effects~\cite{Ciappina_JB,Ciappina_JB2}.

\begin{table}
\begin{center}
\begin{tabular}{|c |c |c |} \hline
Isomer&${ R}_{\rm av}($a.u.$)$&$Q$\\\hline
Ring&$7.8$& $0.21$\\\hline
Bowl&$6.1$ & $0.16$\\\hline
Fullerene&$3.8$& $0.11$\\\hline
\end{tabular}\caption{Average radius and parameter $Q$ for the three $C_{20}$
 isomers. The laser parameters correspond to an intensity of $5\times 10^{14}
 {\rm W}/{\rm cm}^2$ and a wavelength of $800$nm. For the bowl the radius is taken as the
 maximum distance from the symmetry axis.}\label{tab:q}
\end{center}
\end{table}

Along this paper, different harmonic spectra are studied and related to the geometric and
electronic structure of each molecule. Multi-center interference can
give important information about the localization of the atoms, as have been
shown in Ref.~\cite{Ciappina_JB}. Certain modulations along the
harmonic plateau are direct consequence of multi-center interference effects and can be
localized around the points where the recombination matrix element vanishes
\cite{Chirila,Ciappina_JB}. This relation can be explored in order to
analyze and compare the interference effects for different molecular geometries. Other interference patterns are not directly related to the
multi-center nature of the molecule and can be used in order to understand the electron dynamics during HHG.

\section{Theory}
\label{sec:theory}
Along this Section we present the main formulas used in the time-dependent dipole moment in the Lewenstein model. The importance of the molecular recombination matrix element and its calculation is shown in Sec. \ref{sec:rme}.

\subsection{Time-dependent dipole moment in the Lewenstein model}
Let us consider an electromagnetic wave interacting with a molecule or atom. We
assume that the \textit{single-active-electron approximation} (SAE) holds, and just one
electron is ionized. The
Hamiltonian in the length gauge is

\begin{equation}
\hat{H}=\underbrace{\frac{1}{2}\hat{{\bm p}}^2+V(\hat{{\bm r}})}_{\hat{H}_0}\underbrace{\!\!\frac{}{}-e\bm{\mathcal{ E}}(t)\cdot\hat{{\bm r}}}_{\hat{H}_I}\,,
\end{equation}
where $\hat {H}_0$ represents the ground state Hamiltonian, $\hat{H}_I$ is the
interaction with the electromagnetic wave, $V(\hat{{\bm r}})$ represents the Coulomb potential energy, and ${\bm{\mathcal E}}(t)$ is the electric
field that describes the laser interaction. 

The total time-dependent dipole moment can be approximated as~\cite{bib:wbecker}
\begin{equation}
{\bm d}_L(t)\approx -{\rm i}\int_{-\infty}^{t}{\rm d}t'\langle \psi_0(t)|(e\hat{\bm r})\hat U(t,t')\hat H_I(t')|\psi_0(t')\rangle+c.c.\,, \label{eq:dipole}
\end{equation}
where $|\psi_0 (t')\rangle=|\psi_0\rangle e^{-{\rm i}E_0t'}$, is the
molecular ground state of energy $E_0$ at a time $t'$. According to the Lewenstein model, the total evolution operator $\hat U(t,t')$ is
substituted by a Volkov type evolution operator $U^V(t,t')$ which, in the
length gauge, reads
\begin{eqnarray}
U^V(t,t')&=&\int{\rm d}{\bm p}|{\bm p}-e{\bm A}(t)\rangle \langle {\bm 
 p}-e{\bm A}(t')|\nonumber\\
&\times& \exp{\left[-\frac{{\rm i}}{2}\int_{t'}^t ({\bm p}-e{\bm A}(\sigma))^2{\rm d}\sigma\right]}\,.\label{eq:volkov_evolution}
\end{eqnarray}
Here, ${\bm A}(t)$ is the vector potential associated to the oscillating electric field, $\bm{\mathcal{ E}}(t)=-\frac{\partial {\bm A}(t)}{\partial t}$. 

By replacing the total evolution operator by $U^V(t,t')$ in Eq.~\ref{eq:dipole}, the time-dependent dipole moment can be expressed in terms of the \textit{recombination} and \textit{ionization matrix elements} (RME and IME, respectively) as
\begin{eqnarray}
{\bm d}(t)&=&{\rm i}e^2\int {\rm d}{\bm p}\, \int_{-\infty}^{t}{\rm d}t'\exp{\left[-{\rm i} S({\bm p},t,t')\right]}\label{eq:time_dep_dip_mom}\\
&\times&{{\bm d}}_{rec}^*({\bm p}-e{\bm A}(t))\big[{\bm d}_{ion}({\bm p}-e{\bm A}(t'))\cdot \bm{\mathcal{E}}(t')\big]+c.c.,\nonumber
\end{eqnarray}
where the semiclassical action $S({\bm p},t,t')$ is given by
\begin{equation}
S({\bm p},t,t')=\int_{t'}^{t}{\rm d}\sigma\left[\frac{ ({\bm p}-e{\bm A}(\sigma))^2}{2m}+I_p\right]\,. \label{eq:the_action}
\end{equation}
Here, $I_p=-E_0$ is the atomic or molecular ionization potential. The recombination and ionization matrix elements in Eq.~\ref{eq:time_dep_dip_mom} are written as (see, e.g., Refs.~\cite{Ciappina_JB,Chirila,bib:wbecker})
\begin{eqnarray}
{\bm d}_{ion}({\bm p}-e{\bm A}(t'))&=&\langle{\bm p}-e{\bm A}(t')|\hat{\bm r}|\psi_0\rangle\,, \label{eq:dipoleion}\\
{{\bm d}}^*_{rec}({\bm p}-e{\bm A}(t))&=&\langle\psi_0|\hat{\bm r}|{\bm p}-e{\bm A}(t)\rangle\,. \label{eq:dipolerec}
\end{eqnarray}

The molecular ground state $|\psi_0\rangle$ cannot be obtained analytically
for large molecules and it is necessary to approximate it by means of a \textit{linear combination of atomic orbitals} (LCAO). In the position representation, the electron wave function can be considered as a linear superposition of functions centered at the nuclear locations (which are considered static compared to the fast dynamics of the electrons),
\begin{equation}
\langle {\bm r}|\psi_0\rangle\equiv\psi_0({\bm r})=\sum_{j=1}^{N}\sum_{l=1}^{n} C_{l}\phi_{l}({\bm r}-{\bm R}_j)\,, \label{eq:psio}
\end{equation}
where ${\bm R}_j$ is the position of the $j$-th nucleus, $N$ is the total number of atoms constituting the molecule, and $n$ is the number of atomic orbitals considered in the superposition. In Eq.~\ref{eq:psio}, the parameter $C_l$ depends on the molecular structure and 'weight' of each orbital contribution. The function $\phi_{l}({\bm r})$
represents a superposition of \textit{contracted Gaussian functions}, which are constructed as a sum over \textit{Gaussian primitives} (GPs). In a simplified notation and using Cartesian coordinates, we write that

\begin{equation}
\phi_l({\bm r})= N_l\sum_{k=1}^{k_{max}} \eta_k x^a
y^b z^c e^{-\alpha_k {\bm r}^2}\,, \label{eq:gp}
\end{equation}
where $N_l$ is a normalization constant, $\eta_k$ is the superposition coefficient, whereas $a$, $b$, and $c$ are integers such that $a+b+c=l$. The parameter $l$ is related to the angular momentum quantum number
of the specific atomic orbital. In addition, $k_{max}$ is the number of functions necessary to model the orbital. $\alpha_k$ is the so-called \textit{exponent}, which is closely related to the 'spreading' of the Gaussian orbital in the molecule. For larger $\alpha_k$ values, the atomic orbital shows a larger probability that the electron will be localized near to the nucleus. All constants in Eqs.~\ref{eq:psio} and \ref{eq:gp} are obtained from
Hartree-Fock methods for the optimized geometry of the molecules in their ground states. The quantum chemistry computational package GAMESS was used for that effect~\cite{gordon}.

Since within LCAO atomic orbitals are centered at each nuclear position, Eqs.~\ref{eq:dipoleion} and
\ref{eq:dipolerec} involve new oscillatory factors of the type
$e^{\pm \left[{\rm i}({\bm p}-e{\bm A})\cdot{\bm R}\right]}$. Such terms have to be included into
the semiclassical action, and reflect the molecular multi-center nature of the
wave function. The addition of those terms makes the
action explicitly dependent on the nuclear coordinates and has important
consequences in the overall HHG process~\cite{Chirila}. The total time-dependent dipole moment for molecular systems reads
\begin{eqnarray}
\!&{\bm d}(t)={\rm i}e^2\sum_{i=1}^N\sum_{j=1}^N\int {\rm d}{\bm p}\, \int_{-\infty}^{t}{\rm d}t' {\rm e}^{-{\rm i} \mathcal{S}({\bm p},t,t',{\bm R}_i,{\bm R}_j)}\label{eq:dipolefinal}\\
\!\!&\!\!\times{\bm d}_{\rm rec2}^*({\bm p}\!-\!e{\bm A}(t),{\bm R}_i)\!\left[{\bm d}_{\rm ion2}({\bm p}\!-\!e{\bm A}(t'),{\bm R}_j)\!\cdot\!\bm{\mathcal{E}}(t')\right]\!+\!c.c., \nonumber
\end{eqnarray}
where he modified semiclassical action reads, 
\begin{eqnarray}
\mathcal{S}&({\bm p},t,t',{\bm R}_i,{\bm R_j})=\int_{t'}^{t}{\rm d}\sigma\left[\frac{ ({\bm p}-e{\bm A}(\sigma))^2}{2}+I_p\right]\label{eq:mod_action}\\
&+{\bm p}\cdot({\bm R}_j-{\bm R}_i)+e{\bm A}(t)\cdot{\bm R}_i-e{\bm A}(t')\cdot{\bm R}_j\,.\nonumber
\end{eqnarray}

Recombination and ionization matrix elements in Eq.~\ref{eq:dipolefinal} differ from the ones presented in Eqs.~\ref{eq:dipoleion} and \ref{eq:dipolerec} due to the fact that the fast oscillatory terms were incorporated into the action.

Eq.~\ref{eq:dipolefinal} includes a double sum over
the total number of atoms $N$. Such expression takes into account two
different contributions: ionization and recombination at the same atomic
center ($i=j$) and ionization and recombination at two different centers
($i\neq j$). The former case gives rise the so-called \textit{direct
 harmonics} contribution and the latter case to the \textit{transfer harmonics} contribution to HHG~\cite{Chirila}.

In order to calculate the time dependent dipole moment, it is necessary to perform the multidimensional integral in Eq.~\ref{eq:dipolefinal}. Integration over momentum can be approximated by means of the saddle-point method due to the highly oscillatory nature of the factor $\exp{\left[-{\rm i} \mathcal{S}\right]}$. For this technique to be applicable, the remaining parts of the integral have to be slow oscillating (or non-oscillatory) functions of momentum. This is only achieved by the introduction of the other oscillatory terms related to recombination and ionization matrix elements into the modified action~\cite{Chirila}, as it was done in Eq.~\ref{eq:mod_action}.

The saddle point in momentum, denoted as ${\bm p}_s$, is obtained from the relation
 \begin{equation}
 {\bm \nabla}_{\bm p}\mathcal{S}({\bm p},t,t',{\bm R}_i,{\bm R_j})={\bm 0}\,. \label{eq:saddlec}
 \end{equation}
From Eqs.~\ref{eq:mod_action} and \ref{eq:saddlec} we obtain that
\begin{equation}
{\bm p}_s=\frac{1}{t-t'}\left[e\int_{t'}^t{\rm d}\sigma {\bm A}(\sigma)+({\bm R}_i-{\bm R}_j)\right]\,.\label{eq:saddle}
\end{equation}

Under the saddle-point approximation, the time-dependent dipole moment becomes~\cite{Lewenstein1,Chirila}

\begin{eqnarray}
{\bm d}(t)&={\rm i}e^2\int_{0}^{\infty}{\rm d}\tau\left(\frac{2\pi}{\epsilon +{\rm i}\tau}\right)^{3/2}\sum_{i=1}^N\sum_{j=1}^N{\rm e}^{-{\rm i} \mathcal{S}({\bm p}_s,t,t-\tau,{\bm R}_i,{\bm R}_j)}\nonumber\\
&\times{{\bm d}}^*_{rec2}({\bm p}_s-e{\bm A}(t),{\bm R}_i)\nonumber\\
&\times\big[{\bm d}_{ion2}({\bm p}_s-e{\bm A}(t-\tau),{\bm R}_j)\cdot\bm{ \mathcal{E}}(t-\tau)\big]+c.c.\,, \label{eq:finaldipoletime}
\end{eqnarray}
where $\epsilon$ is an infinitesimal regularization constant and $\tau=t-t'$ is the so-called \textit{return time}~\cite{Lewenstein1}. The integration over $\tau$ is done numerically.

Let us note that Eq.~\ref{eq:saddle} has important consequences. In the first place, if the semiclassical action was not modified, the saddle point in momentum would always be parallel to the laser polarization direction. The inclusion of the term $\frac{1}{\tau}({\bm R}_i-{\bm R}_j)$, when $i\neq j$, guarantees that the electron has a momentum component perpendicular to the driving laser field. This new term contributes, in a very important way, to the ellipticity of the resulting harmonics, even if they are originated from linearly polarized radiation~\cite{Etches2}. The polarization of the resulting harmonics depends strongly on the molecular orientation, as it was experimentally observed in diatomic molecules (see, e.g.,~\cite{Zhou2}).

\subsection{Recombination matrix element}
\label{sec:rme}
As it was mentioned before, harmonic responses from molecules can exhibit strong modulations of peaks intensity along the plateau. In general, when the atomic 
case is considered, the returning electron wave packet collides with a unique center whereas, in the molecular case, many atomic centers are present. This results in pronounced interference modulations of the spectral response. The position and intensity of the suppressed harmonics depend strongly on the molecular orientation~\cite{Lein} and on the orbital symmetry. As the interference effects are directly related to the geometric distribution of the atomic centers, a detailed analysis of the minima can provide valuable information about the molecule. It has been proven that, for other fullerenes and multi-atomic systems, some of the modulations along the plateau are strongly related to the recombination matrix element values, presenting minima in the region when the RME vanishes~\cite{Ciappina_JB,Ciappina_JB2}. Other type of local spectral modulations is traditionally related to interferences between different electron trajectories~\cite{bib:lein2,Lewenstein1}.

The molecular RME, as a function of the kinetic momentum of the electron, ${\bm \Pi}({\bm p},t)={\bm p}-e{\bm A}(t)\equiv {\bm \Pi}$, is given by
\begin{eqnarray}
&{{\bm d}}^*_{rec}({\bm \Pi})=\frac{1}{(2\pi)^{3/2}}\sum_{i=1}^{N}\sum_{l=0}^n C_l N_l\sum_k\eta_k\label{eq43}\sum_{a,b,c} \int {\rm d}{\bm r}\nonumber\\
&\times (x-x_i)^a(y-y_i)^b (z-z_i)^c{\rm e}^{-\alpha_k ({\bm r}-{\bm R}_i)^2}{\bm r}{\rm e}^{{\rm i}{\bm \Pi}\cdot{\bm r}}\,. 
\end{eqnarray}
It is evident that the RME values depend on the nuclear positions ${\bm R}_i$ and the set of parameters $C_l$, $N_l$, $\eta_k$, and 
$\alpha_k$. As the RME is a function of the kinetic momentum of the electron during recombination and the harmonic response depends on the frequency of the emitted photon, it is necessary to relate those two quantities by means of the energy conservation equation~\cite{Ciappina_JB,Ciappina_JB2},
\begin{equation}
\omega=\frac{1}{2}{\bm \Pi}^2+I_p\,. \label{eq:pi}
\end{equation}
where $\omega$ is the frequency of the harmonic photon. Eq.~\ref{eq:pi} allows us to find the harmonic frequencies for which the RME vanishes.

As it has been pointed out before, when a multi-center system is considered, the saddle point in momentum can have non-zero components in all directions 
[see, Eq.~\ref{eq:saddle}] depending on the molecular geometry and orientation. If the internuclear distance is small enough (more precisely, 
if the parameter $Q$ [Eq.~\ref{eq:q}] is small enough), it is expected that the main component of the saddle point in momentum is parallel to the laser polarization direction. 
The other components can be fairly large as well, especially for small $\tau$ values. As the first approximation, the RME is calculated as a function of the harmonic 
order by increasing the kinetic momentum ${\bm \Pi}$ in just one direction and setting the other components as constant. This gives an idea of the approximated behavior of the RME along the plateau, accounting for the three momentum components separately.

\section{Results}

The goal of this Section is to analyze the relation between the geometric distribution of atomic centers, molecular symmetry and orientation of different $C_{20}$ isomers 
with the spectral properties of the harmonic response. Modulations of peak intensities along the plateau, together with the polarization properties of the emitted radiation, are going to be explored in order to gain a better understanding of the overall HHG process.

In our calculations, the time-dependent dipole moment is calculated by performing the numerical integration over $\tau$ in Eq.~\ref{eq:finaldipoletime}. The dipole acceleration in frequency domain is obtained from the Fourier transform of each one of the time-dependent dipole moment vector components . The laser field is described as a semi-infinite and monochromatic plane wave, with an oscillating electric field given by the relation

It is considered that the laser field can be polarized along the $x$-, $y$-, or $z$-directions, whereas the wavelength and intensity are the same as described in Sec.~\ref{sec:theory}. In order to obtain an expression for the molecular
ground state, the LCAO, with coefficients generated by standard quantum chemistry software, was used. 

A linearly-polarized laser field interacting with large, non-linear molecules can
generate important harmonic responses in directions perpendicular to the driving field polarization. Thus, the complete study of the
HHG should involve three different dipole components [$d_x(t)$, $d_y(t)$, and $d_z(t)$] for each linear polarization. In total, nine responses are
expected from each isomer. Every harmonic response is going to be denoted as $d^{(i)}_{j}$ ($i,j=x,y,z$), where the bottom index denotes the harmonic polarization whereas the top index relates to the driving field polarization direction with respect to the coordinates and orientation of the systems shown in Fig. \ref{fig:all}. 

\subsection{Harmonic spectra from the HOMO}

In this Section, the harmonic responses from the HOMO of the ring, bowl, and cage are going to be analyzed. In Sec.~\ref{sec:multiorbital}, the influence of the HOMO-1 and inner molecular orbitals on the harmonic signal and RME values is going to be explored.

\subsubsection{Results for $C_{20}$ fullerene}
\label{section:cage}

\begin{figure}
\includegraphics[width=0.9\columnwidth]{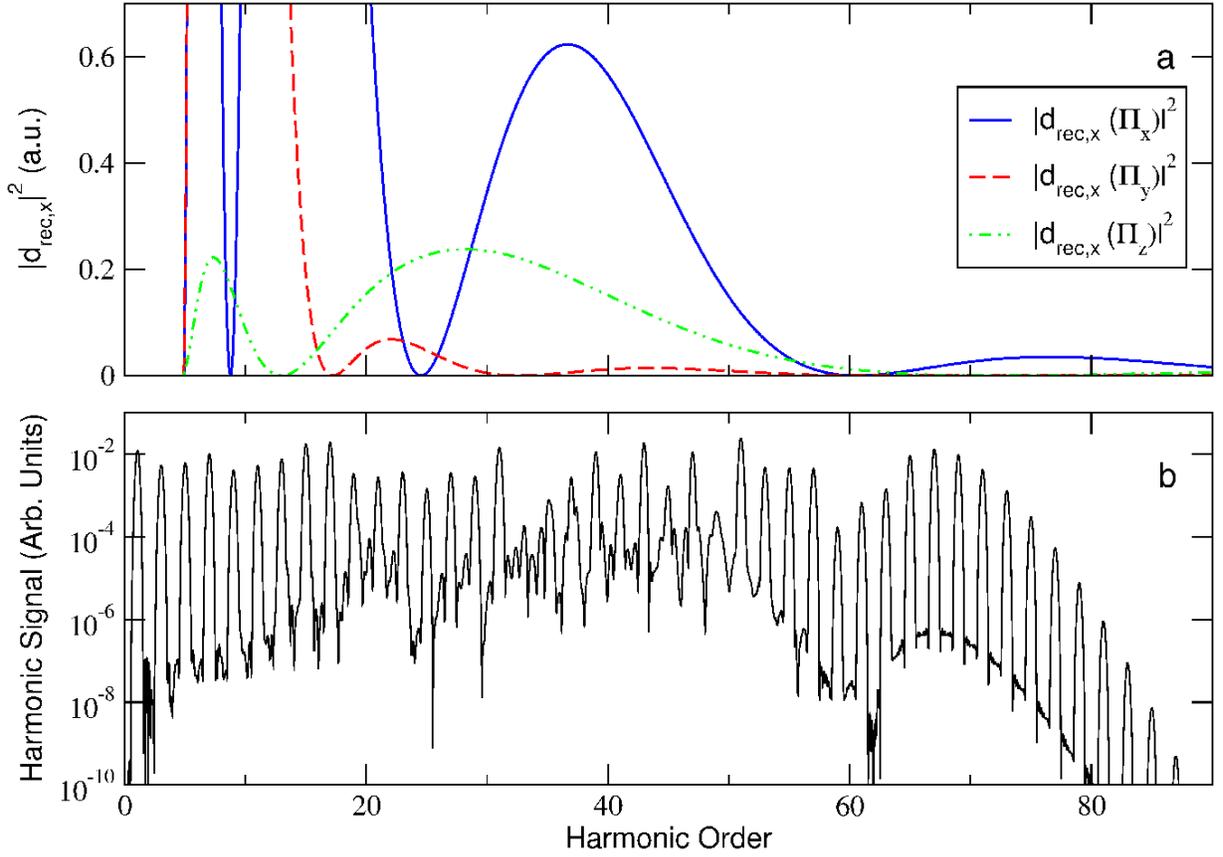}
\caption{Panel (a) presents the {modulus squared} of the recombination matrix element $x$-component, calculated according to Eq.~\ref{eq43}, as a function of $\Pi_x$, $\Pi_y$, and $\Pi_z$ for the $C_{20}$
 fullerene. Panel (b) shows the harmonic response $d^{(x)}_{x}$ from the same structure. The electric field is described by Eq.~\ref{eq:semiinfinite_sine} and it is considered to be polarized along the $x$-direction. The wavelength corresponds to $800$nm and the intensity is $I=5\times 10^{14}{\rm W}/{\rm cm}^2$. }
\label{fig:C20f5_Ex}
\end{figure}

When the laser field polarized along the $x$-axis interacts with the symmetric cage [Fig.~\ref{fig:all}(c)], a strong harmonic response is observed in the $x$-direction, with a plateau
characterized by multiple modulations in peak intensities. The spectral response is presented in panel (b) of Fig.~\ref{fig:C20f5_Ex}. The most visible minima
are located in the regions between the 23th and 27th, 31st and 39th, and 55th and 65th harmonic orders (HO). 

In panel (a), we present the modulus {squared} of the recombination matrix element's $x$-component, calculated according to Eq.~\ref{eq43}. As the momentum of the electron is not necessarily parallel to the laser field polarization, 

$\left|d_{\rm rec,x}({\bm \Pi})\right|^2$ is shown as a function of $\Pi_x$, $\Pi_y$,
and $\Pi_z$, separately. As it was mentioned before, it is expected that, for molecules with small $Q$ values, the major
contribution from the saddle points in momentum [Eq.~\ref{eq:saddle}] should be
parallel to the laser polarization (solid blue line). It can be seen that the spectral minima from the 23th to 27th and from the 55th to
65th harmonic orders (the most prominent modulations) match very well with the zeroes of the RME being the function of $\Pi_x$. The region from the 19th to 21st (less pronounced) and
from the 31st to 39th harmonics are located near the points where the RME as a function of $\Pi_y$ vanishes (dashed red line). 

The most pronounced minima agree with the zeroes of the RME and, therefore, one can attribute those modulations to interference effects related to the multi-center nature of the molecule. Similar minima have been observed in other harmonic responses obtained from larger icosahedral fullerenes, and have been attributed to the same cause~\cite{Ciappina_JB}.

\begin{figure}
\includegraphics[width=0.75\columnwidth]{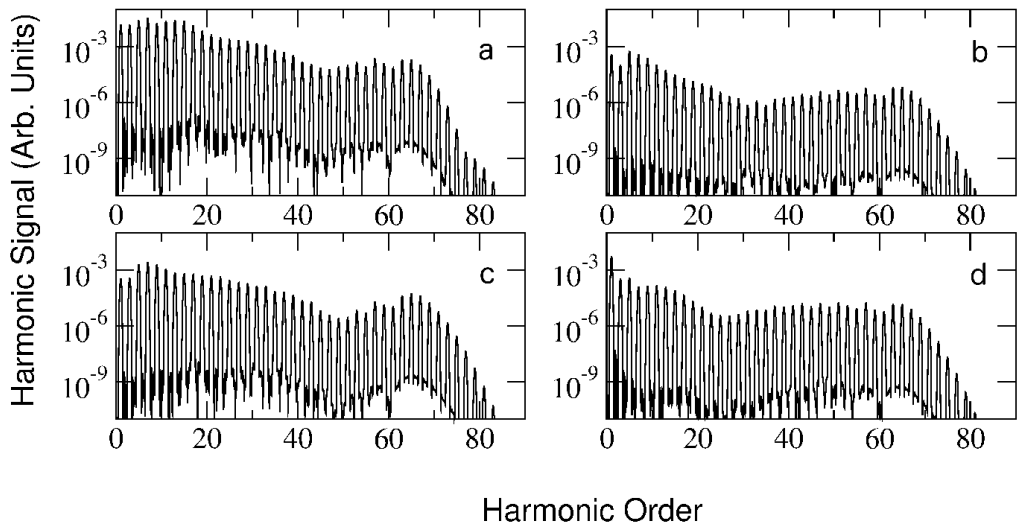}
\caption{The left column presents the harmonic response from the $C_{20}$ fullerene when the laser field is polarized along the $y$-direction [the $d^{(y)}_{y}$ harmonic response is presented in panel (a) and $d^{(y)}_{z}$ response in panel (c)]. The right column corresponds to a polarization of the driving field along the $z$-direction  [{$d^{(z)}_{y}$} response is presented in panel (b) and $d^{(z)}_{z}$ response in panel (d)]. The remaining laser field parameters are the same as in Fig.~\ref{fig:C20f5_Ex}.}
\label{fig:C20f5_tot}
\end{figure}

It is worth noting that, when the electric field is polarized
along the $x$-direction for the given molecular orientation, the only
harmonic response is obtained parallel to it (i.e., the responses $d^{(x)}_{y}$ and $d^{(x)}_{z}$ are suppressed).
If the driving field is polarized along the $y$- or $z$-directions, the harmonic signal shows a different behavior (Fig.~\ref{fig:C20f5_tot}). In the first place, for each laser polarization we obtain two responses: one
parallel to the electric field ($d^{(y)}_{y}$ and $d^{(z)}_{z}$) and one perpendicular to it ($d^{(z)}_{y}$ and $d^{(y)}_{z}$). Harmonics in the $x$-direction are completely suppressed.

It can be seen from Fig.~\ref{fig:C20f5_tot} that, in the plateau region, all spectra present smoother variations of the peak intensities as compared to the $d^{(x)}_{x}$ case (i.e., the envelope of the peaks shows less modulations). In those cases, the total RME $y$- and $z$-components vanish independently of the momentum direction. This can be considered as a consequence of symmetry of the molecular orbitals.  As we have checked, if any of the atoms is artificially displaced from its original position the two RME components present strong oscillations, similar to the $d^{(x)}_{x}$ case. The very smooth modulations along the plateau are, as it will be shown later, consequences of interferences between quantum trajectories.

Up to now, the analysis of modulations of the peak intensities along the HHG plateau from three different $C_{20}$ structures has been based on the importance of the RME. It is clear that this quantity contains important information about the molecular configuration and it is related to multi-center interference effects. Another factor which provides substantial information about the process is the modified  semiclassical action, as it will become clear along this Section. 

The relation between multi-center interference effects 
in diatomic molecules and its harmonic response was 
originally proposed by Lein et al. in Refs.~\cite{Lein,bib:lein2}, but other interference features of different nature were ignored. The direct observation of  interference between quantum trajectories (or the quantum path interference (QPI) phenomenon) was originally proposed for a single atom~\cite{bib:zair}, where  multi-center effects play no role. Intuitively, the QPI can be understood by analyzing the Lewenstein model: the electrons which contribute to HHG can follow two paths, one short and one long. During the excursion to the continuum the electron wave packet spreads and acquires a phase, given by the semiclassical action, which depends on the ionization and recombination times. When the recollision takes place, electrons with different phases interfere. This has direct consequences on the harmonic spectrum. The inclusion of QPI to the analysis of HHG from diatomic molecules has been recently studied by Yang et al. (see, Ref.~\cite{bib:yang}) and has proven to modify in a very important way the general behavior of the harmonic plateau. New minima were observed with no relation to the multi-center interference but they were unequivocally 
related to QPI processes, which follows from the time-frequency analysis. Such effects have shown to be important even for ultrashort driving pulses~\cite{bib:yang}.
\begin{figure}
\includegraphics[width=0.8\columnwidth]{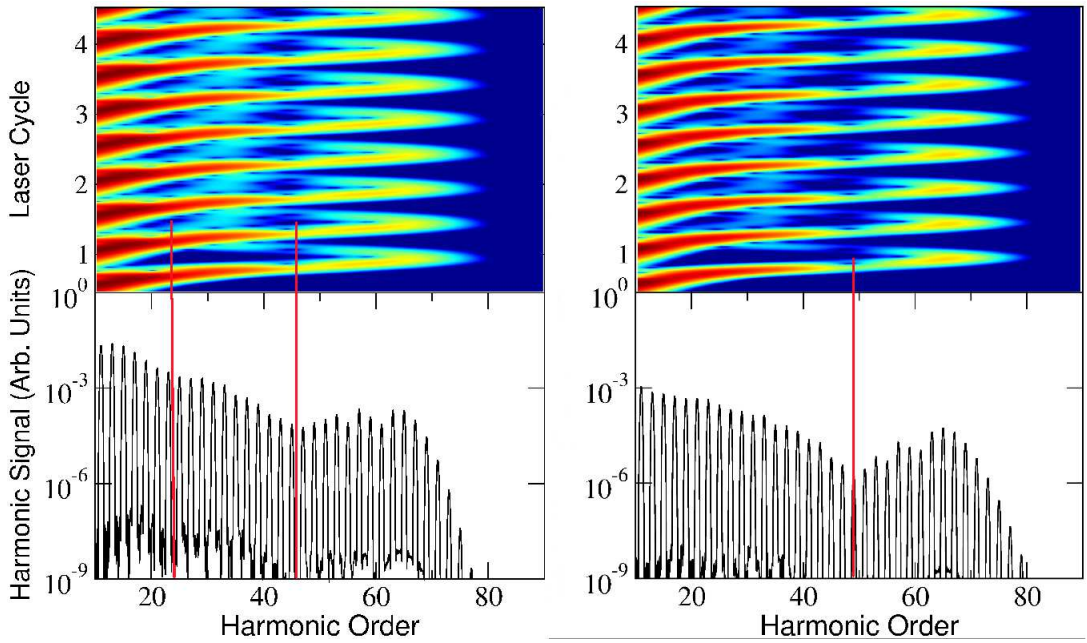}
\caption{Left panel shows the time-frequency analysis (top) for the $d^{(y)}_{y}$ harmonic response from the cage and the corresponding HHG spectrum is presented in lower figure. The laser parameters are {the same} as in Fig.~\ref{fig:C20f5_tot}. The first vertical red line shows the starting point of the region where the interference between quantum trajectories begins to be evident. The second vertical line points at the region where the destructive interference is maximal. Right panel presents the time-frequency analysis of the response $d^{(y)}_{z}$ from the cage (top), which is shown in lower pane. The vertical line points at the minimum of the harmonic spectrum envelope, which can be directly related to QPI effects.}
\label{fig:Eyy_Eyz_wavelet}
\end{figure}

The QPI analysis in the atomic case is straightforward as just the long and short electron paths are considered. If a multi-center system is taken into account, the complexity of the problem increases with the number of atoms. It is evident from Eq.~\ref{eq:mod_action} that the modified semiclassical action does not depend  only on the ionization and recombination times, but depends as well on each of the nuclear coordinates. In this sense, short and long trajectories are composed of \textit{direct} and \textit{transfer} trajectories which may interfere and generate very rich harmonic responses.

\begin{figure}
\includegraphics[width=0.8\columnwidth]{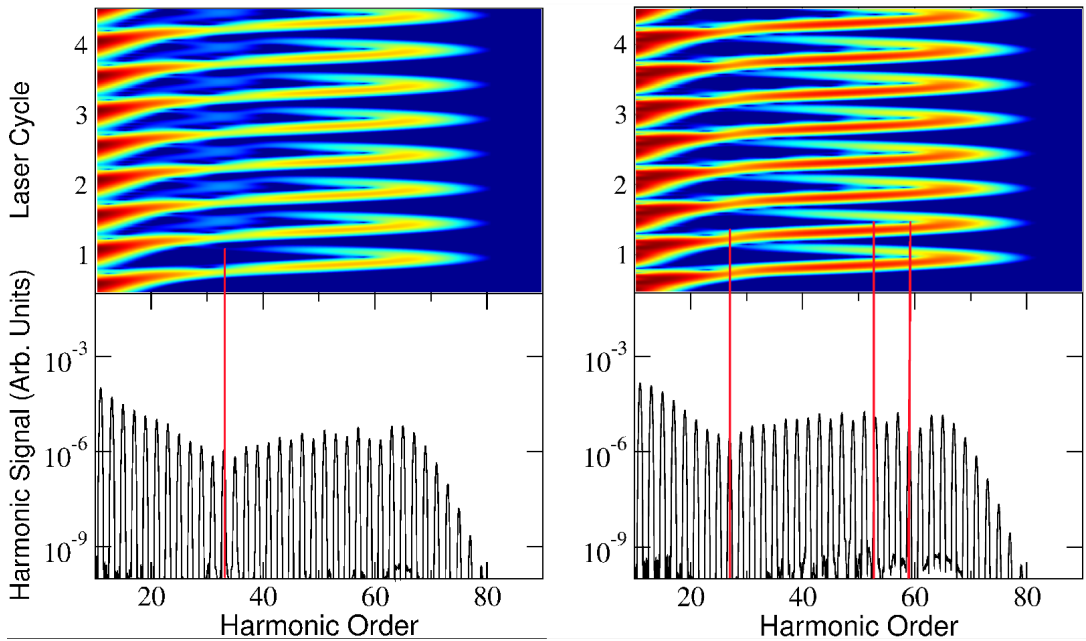}
\caption{The same as in Fig.~\ref{fig:Eyy_Eyz_wavelet} but for the time-frequency analysis of the $d^{(z)}_{y}$ [panel (a)] and $d^{(z)}_{z}$ [panel (b)] harmonic responses [panels (c) and (d), respectively]. The red vertical lines point at the regions where the modulations in the spectral response coincide with quantum trajectories interference. }
\label{fig:Ezy_Ezz_wavelet}
\end{figure}

In Sec. \ref{section:cage}, the HHG spectra from $C_{20}$ fullerene have been extensively studied by interpreting the modulations of the peak intensities along the  plateau as multi-center interferences, but some of such modulations cannot be matched with minima of the RME. It has been proven that the harmonic responses related  to oscillating RMEs exhibit heavily modulated plateaus and many of the minima match fairly good with the zeroes of the RME. The harmonic responses related to non-oscillating 
RMEs present smooth plateaus with soft modulations of the peak intensities, e.g., the $d^{(y)}_{y}$, $d^{(y)}_{z}$, $d^{(z)}_{y}$, and $d^{(z)}_{z}$ spectra obtained from the fullerene. In order to identify the nature of such soft variations, the time-frequency analysis was performed for the four aforementioned responses and it is presented together with the respective spectra in Figs.~\ref{fig:Eyy_Eyz_wavelet} and~\ref{fig:Ezy_Ezz_wavelet}. In general, all plots show the well defined long and short trajectories with interference effects at certain frequencies. Those interferences can be observed as bifurcations or interactions between the otherwise clear, well defined, and independent paths. The vertical red  lines identify the position of the spectral minima or regions which present modulations that can be related to QPI in the time-frequency analysis. 
Starting with Fig.~\ref{fig:Eyy_Eyz_wavelet}, in panel (a)(top) one can see that the $d^{(y)}_{y}$ response exhibits interference features from around the 20th harmonic order, where a decrease of the peak intensities can be observed. The most evident destructive interference effects are present between the 40th and 50th harmonic orders,  where the minimum of the plateau is present [lower figure]. The $d^{(y)}_{z}$ response shows QPI effects starting from roughly the 30th harmonic, when the intensity of the peaks starts to decrease [see, panels (b) and (d) of Fig.~\ref{fig:Eyy_Eyz_wavelet}]. The most pronounced destructive interference effect is clearly located between the 45th and 55th harmonic orders, which again coincides with the position of the minimum. The time-frequency analysis shows much less pronounced (short)(long?) trajectories. 

When the laser field is polarized along the $z$-axis, fewer interference effects are observed. Particularly interesting is the $d^{(z)}_{y}$ time-frequency analysis [Fig.~\ref{fig:Ezy_Ezz_wavelet}, panel (a), left panel top figure]. It exhibits very well defined short and long trajectories except for the region of the 25th-40th harmonics, where the long trajectories are clearly distorted and less pronounced as a consequence of strong interference. This path interference coincides with exact position of the broad spectral minimum. Finally, the $d^{(z)}_{z}$ case [Fig.~\ref{fig:Ezy_Ezz_wavelet} panel (b), right panel] presents relatively small QPIs, being the most intense at around the 25th-35th,53rd and 59th harmonics, when the two trajectories seem to interfere the most and the minimum of the plateau is present. There are other destructive interactions near to the cutoff.

In closing this Section, it is worth noting that the smooth variations of the spectral response for the cases where the multi-center effects are less important can be directly related to QPI effects (i.e., to the acquired phase of the electron and to the modified semiclassical action).

\subsubsection{Results for $C_{20}$ ring}

\begin{figure}
\includegraphics[width=0.9\columnwidth]{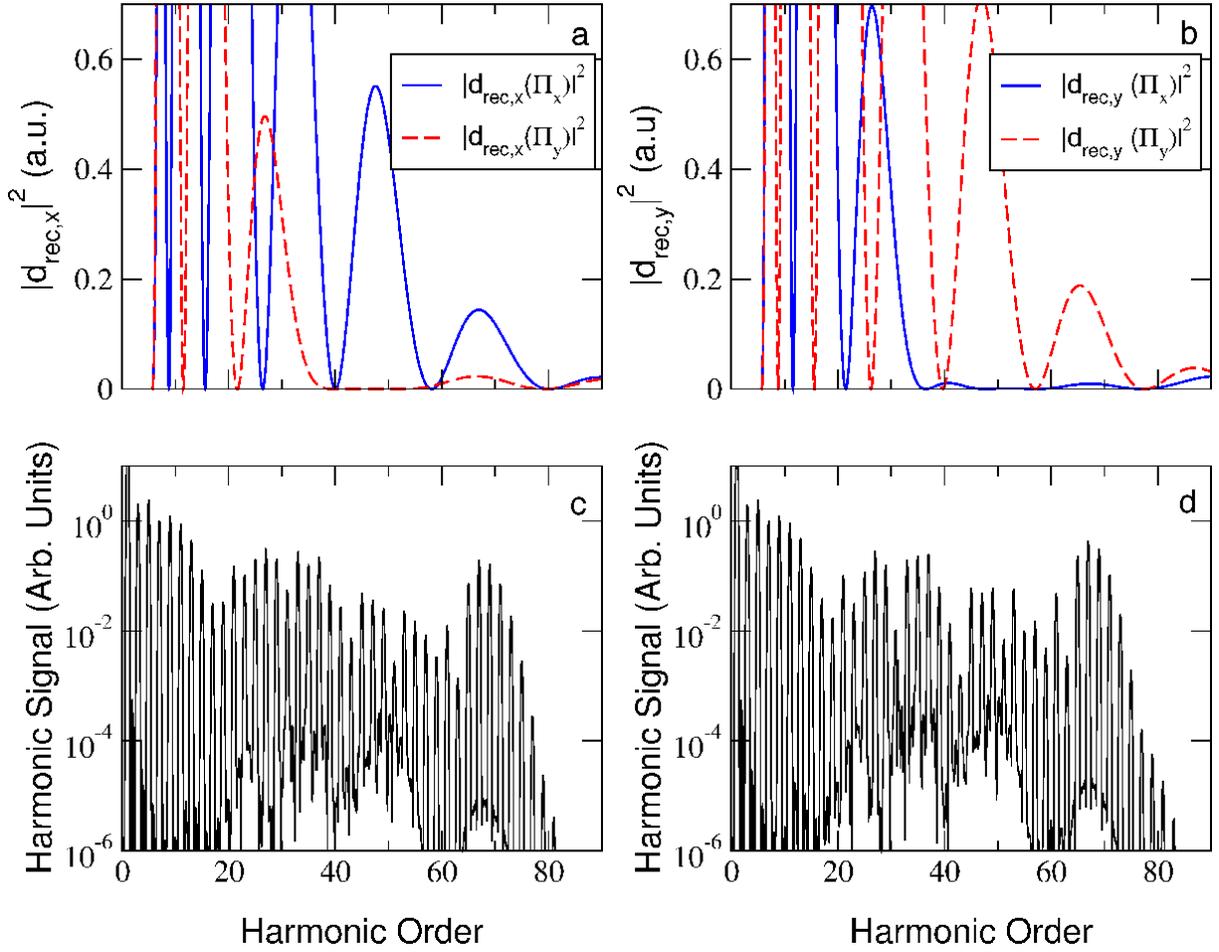}
\caption{Panel (a) presents the modulus {squared} of the recombination matrix element $x$-component as a function of either $\Pi_x$ (solid blue line) or $\Pi_y$ (dashed red line) for the ring. The RME as a function of $\Pi_z$ vanishes. Panel (b) shows the same but for the $y$-component of the RME. In panels (c) and (d) the harmonic responses $d^{(x)}_{x}$ and $d^{(y)}_{y}$ are shown, respectively. The remaining laser parameters are the same as in Fig.~\ref{fig:C20f5_Ex}.}
\label{fig:C20f2_rec_mat}
\end{figure}

The ring [Fig.~\ref{fig:all}(a)] is an interesting $C_{20}$ isomer due to its planar configuration and its highly symmetric structure with respect to the $z$-axis. When the $d^{(x)}_{x}$ and $d^{(y)}_{y}$ responses are considered, strongly modulated harmonic plateaus are observed [panels (c) and (d) of Fig.~\ref{fig:C20f2_rec_mat}]. Similarly to the case of the $C_{20}$ fullerene, the modulations of the HHG spectra are related to oscillating components of the RME [with their modulus {squared} presented in panels (a) and (b) of the same figure]. Some of the most pronounced minima (located from the 37th to 47th and from the 51st to 65th HOs) can be directly related to the zeroes of the RME components with momentum parallel to the polarization axis [solid blue line in panel (a) for $d^{(x)}_{x}$ and red dashed line in panel (b) for $d^{(y)}_{y}$]. In panels (c) and (d) of Fig.~\ref{fig:C20f2_rec_mat}, the peaks in the range between the 17th and 23rd  harmonics present strong modulations. In this region the RME components reach zero at several points.

\begin{figure}
\includegraphics[width=0.75\columnwidth]{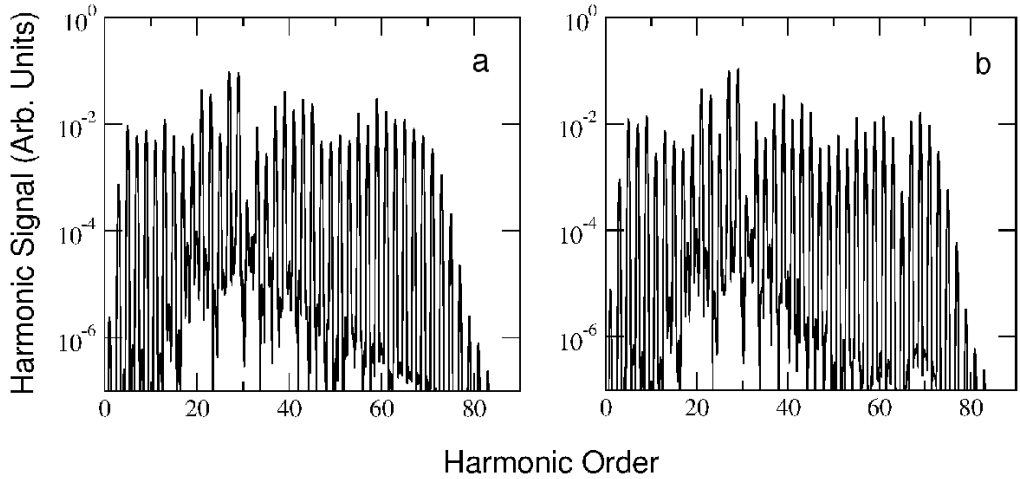}
\caption{HHG spectra generated from the ring. The harmonic responses correspond to (a) {$d^{(x)}_{y}$ and (b) $d^{(y)}_{x}$}. The laser parameters are the same as in Fig.~\ref{fig:C20f5_Ex}. }
\label{fig:C20f2_tot}
\end{figure}

When the laser field is polarized along the $x$- or $y$-directions, two other harmonic responses are observed ($d^{(x)}_{y}$ and $d^{(y)}_{x}$), which are presented in Fig.~\ref{fig:C20f2_tot}. Due to the highly symmetric structure (wave function and geometry of the molecule) in the $xy$-plane, one can expect to obtain similar results for both laser polarization along $x$- and $y$-directions.
 
This is confirmed by our results, as one can see by comparing panel (a) with panel (b) in Fig.~\ref{fig:C20f2_tot}, and panels (c) and (d) in Fig.~\ref{fig:C20f2_rec_mat}. There are relatively small differences in the intensity of some peaks, which are related to imperfect symmetry of the wave function introduced by the numerical error in the ab-initio calculations using the GAMESS code. Nevertheless,  the overall spectral trend shows similar characteristics.

When the driving field is polarized along the $z$-axis, just one harmonic response, $d^{(z)}_{z}$, is observed [see, Fig.~\ref{fig:C20f2_dzz} (c)]. In this case, the $z$-component of the RME vanishes for all components of momentum, and the resulting harmonic spectrum presents the characteristic smooth plateau.

It is worth noting that, for the ring, all the RME components as a function of $\Pi_z$ vanish. It is not expected that this particular momentum direction would contribute to the harmonic response when the laser field is polarized along the $x$- or $y$- directions. This is due to the planar configuration of the isomer and the fact that its symmetry axis is aligned with the $z$-direction. By inspecting Eq.~\ref{eq:saddle} one can clearly see that the $z$-component of the saddle point in momentum is zero.

\subsubsection{Results for the $C_{20}$ bowl}

\begin{figure}
\includegraphics[width=0.8\columnwidth]{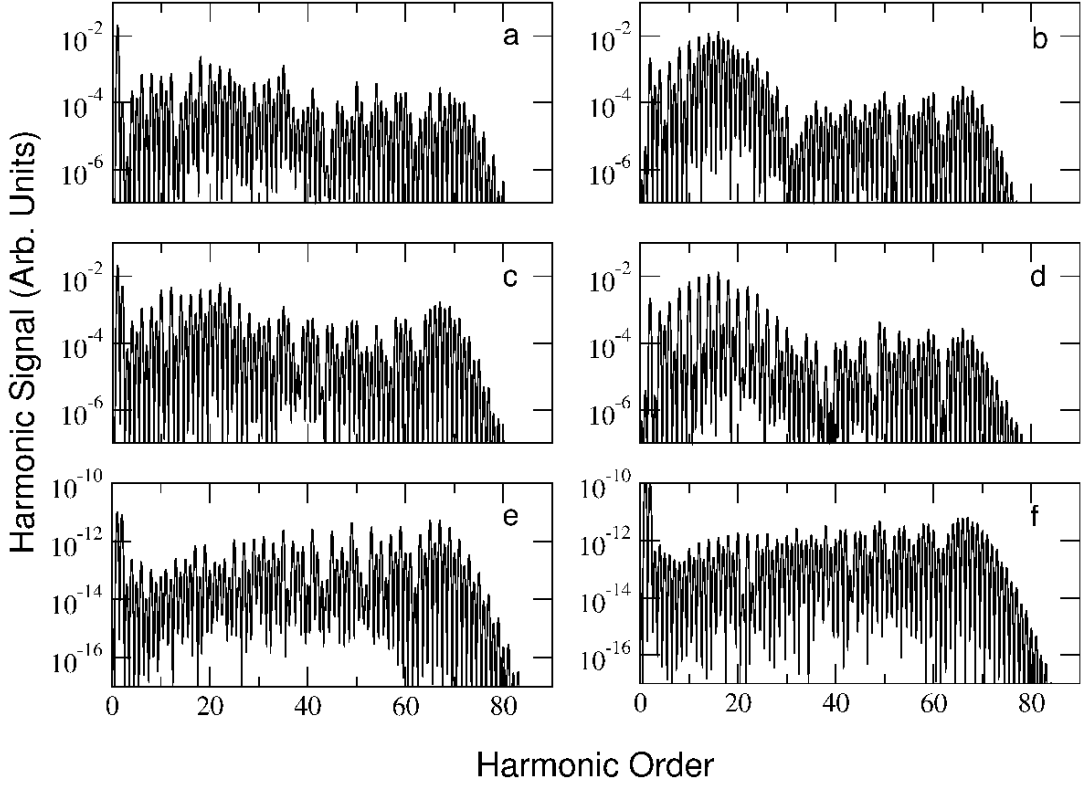}
\caption{HHG spectra generated from the bowl. The electric field is polarized along the $x$-axis (top row), along the $y$-axis (middle row), and along the $z$-axis  (bottom row). Harmonic responses are (a) $d^{(x)}_{y}$, (b) $d^{(x)}_{z}$, (c) $d^{(y)}_{x}$, (d) $d^{(y)}_{z}$, (e) $d^{(z)}_{x}$, and (f) $d^{(z)}_{y}$.}
\label{fig:C20f3_tot}
\end{figure}

\begin{figure}
\includegraphics[width=0.8\columnwidth]{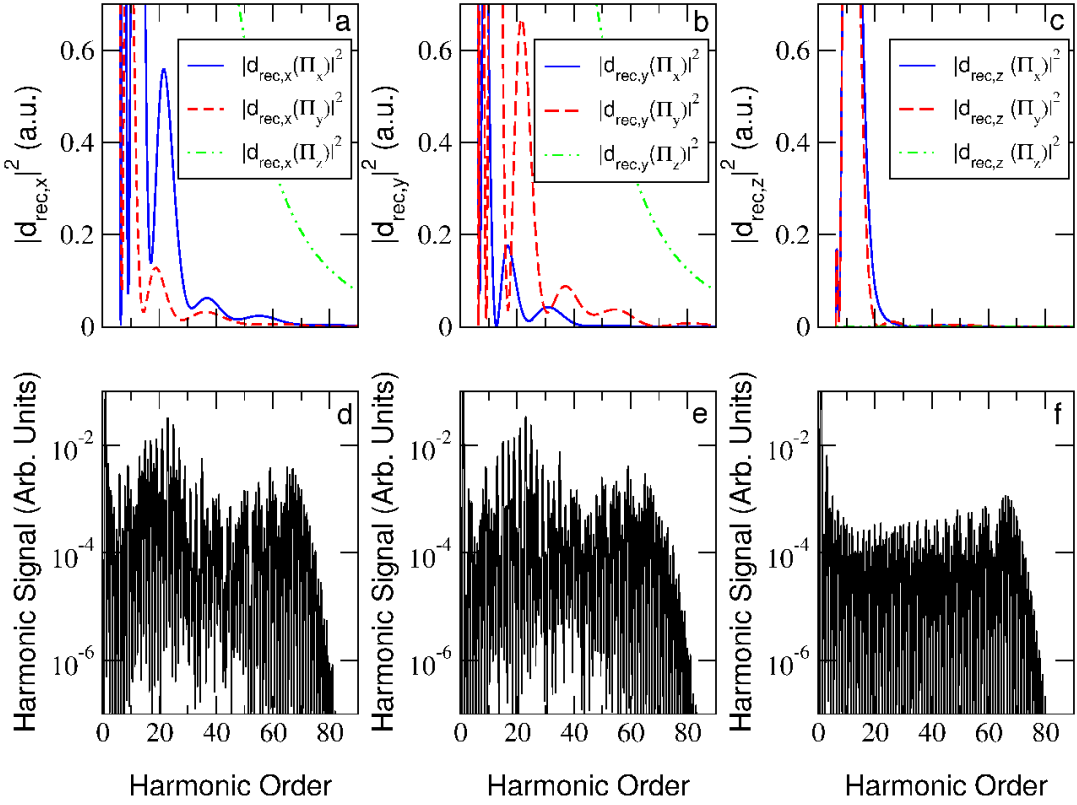}
\caption{(a) Modulus {squared} of the RME $x$-component as a function of $\Pi_x$, $\Pi_y$, and $\Pi_z$ calculated for the bowl. (b) The same as in panel (a) but for the $y$-component of the RME. (c) The same as in panel (a) and but for the $z$-component of the RME. (d) Harmonic response $d^{(x)}_{x}$. (e) Harmonic response $d^{(y)}_{y}$. (f) Harmonic response $d^{(z)}_{z}$. }
\label{fig:C20f3_rec_mat}
\end{figure}

The bowl [Fig.~\ref{fig:all}(b)] is the least symmetric of the studied $C_{20}$ isomers, which directly influences the harmonic spectrum. In the first place, all different combinations of laser field and harmonics polarizations are nonzero (Figs.~\ref{fig:C20f3_tot} and \ref{fig:C20f3_rec_mat}). As in the previous cases, the position of the cutoff coincides with the $3.17 U_p+I_p$ rule and no nonphysical extension of the plateau is observed. The atomic distribution of this molecule generates a static dipole moment pointing along the $z$-direction, as reported in Table \ref{tab:ionization_potentials}. It has been shown that the harmonic responses from polar molecules show the presence of even and odd harmonics, which is related to the lack of inversion symmetry (see, e.g. ,~\cite{bib:gavrilenko} and references therein). All responses obtained from the bowl present strong even and odd harmonics.

All spectra in Figs.~\ref{fig:C20f3_tot} and~\ref{fig:C20f3_rec_mat} exhibit several modulations along the plateau, being the less prominent for the $d^{(z)}_{z}$ case. In Fig.~\ref{fig:C20f3_rec_mat}, the $d^{(x)}_{x}$, $d^{(y)}_{y}$ and $d^{(z)}_{z}$ spectra are plotted together with the modulus squared of the three components of the RME as a function of $\Pi_x$, $\Pi_y$, and $\Pi_z$. For this particular structure, due to the generally complicated shape of the plateau, a comparison between RME's zeroes and expected minima of the spectral envelope is not as clear as in the two previous cases. When analyzing the $d^{(x)}_{x}$ response [Fig.~\ref{fig:C20f3_rec_mat} (d)], a minimum can be found between the 21st and 35th harmonics, which coincides with
minima of the modulus squared of the RME's $x$-component as a function of $\Pi_x$ and $\Pi_y$ (solid blue and dashed red lines in Fig.~\ref{fig:C20f3_rec_mat} (a), respectively). 
Other modulations can be attributed to the combined oscillations of the RME for different momenta. Interference effects between electron quantum trajectories (short and long) starting from different atomic positions are expected to have another important role in the peak intensities modulation. What is certain is that the $\Pi_z$ component does not contribute to destructive interference due to the absence of zeroes in the
RME [dotted-dashed green line in Fig.~\ref{fig:C20f3_rec_mat} (a)]. A similar analysis can be applied to the spectral response $d^{(y)}_{y}$ shown in Fig.~\ref{fig:C20f3_rec_mat} (e). A general decrease of the peaks strength
is present between the 25th and 35th harmonics, which corresponds to a
minimum in the modulus squared of the RME $y$-component associated to $\Pi_y$ (dashed red curve in \ref{fig:C20f3_rec_mat} (b)). The following peaks show lower intensities, which can be attributed to the small RME absolute values.

The $d^{(z)}_{z}$ spectrum (see Fig.~\ref{fig:C20f3_rec_mat} (f)), shows a different behavior. Even though there are present small modulations in the peak intensities along the plateau, the general trend is more uniform than in the previous two cases. It can be explained by the rapid decrease of the RMEs associated with $\Pi_x$ and $\Pi_y$ (solid blue and dashed red curves in Fig.~\ref{fig:C20f3_rec_mat} (c)). The modulus squared of the matrix element as a function of $\Pi_z$ (dotted-dashed green curve) is expected to have a strong influence on the spectrum and it vanishes at the considered frequencies. In the absence of oscillations, the harmonic response presents a relatively smooth plateau.

As it can be seen in all spectra obtained from the bowl, peaks corresponding to even multiples of the driving frequency, in addition to the well defined odd harmonics, are present. This feature is not observed for other two structures, as it is a direct consequence of symmetry of the molecule. Even though the bowl has an axial symmetry,  the structure presents a symmetry breakdown with respect to the $xy$-plane. The appearance of even harmonics in the spectrum has been observed when the Lewenstein model  is applied, in both length and velocity gauges, to atoms or molecules displaced from the origin of coordinates~\cite{Chirila}. The crucial point to observe the appearance 
of even harmonics is the absence of an inversion point of symmetry in the molecule.

It is worth noting that all the obtained harmonic responses from $C_{20}$ isomers present a sharp cutoff at the position described by the relation (\ref{eq:cutoff})
and no unphysical extensions are observed. This is expected due to the small $Q$ values and it agrees with the observations presented in Refs.~\cite{Ciappina_JB,Ciappina_JB2}.

\subsection{Influence of the HOMO-1 orbital}
\label{sec:multiorbital}
\begin{table}
\begin{center}
\begin{tabular}{|c|c|c|}
\hline
Isomer & ${ Cutoff}_{H-1}$ & ${Cutoff}_{H}$ \\\hline
{ Ring } & $66.84$ & $66.81$ \\\hline
{Bowl }& $67.63$ & $67.47$ \\\hline
{Fullerene} & $66.99$ & $66.09$\\\hline
\end{tabular}\caption{Cutoff position for HOMO and HOMO-1 calculated according to Eq.~\ref{eq:cutoff} using the ionization potential values reported in Table \ref{tab:ionization_potentials}. The laser field is approximated as a semi-infinite sinusoidal plane wave with intensity $I=5\times 10^{14} {\rm W}/{\rm cm}^2$  and wavelength $\lambda=800$ nm.}
 \label{tab:cutoff}
\end{center}
\end{table}

 One of the challenges related to studying HHG from large molecules and nanostructures is associated to the fact that the considerable number of electrons can cause that it is relatively easily to ionize the system. Note that the ionization potentials for several valence orbitals can differ by less than one electronovolt (for HOMO and HOMO-1, see, Table \ref{tab:ionization_potentials}). For this reason, it is often necessary to consider multiple {molecular} orbitals in the LCAO expansion. The contributions from different orbitals can add new interference features. 
In this Section, we present how, for the {$C_{20}$ isomers}, the addition of HOMO-1 modifies the overall structure of the HHG spectra.
The RMEs behavior as a function of momentum is expected to change as well.

\begin{figure}
\centering
\includegraphics[width=0.8\columnwidth]
{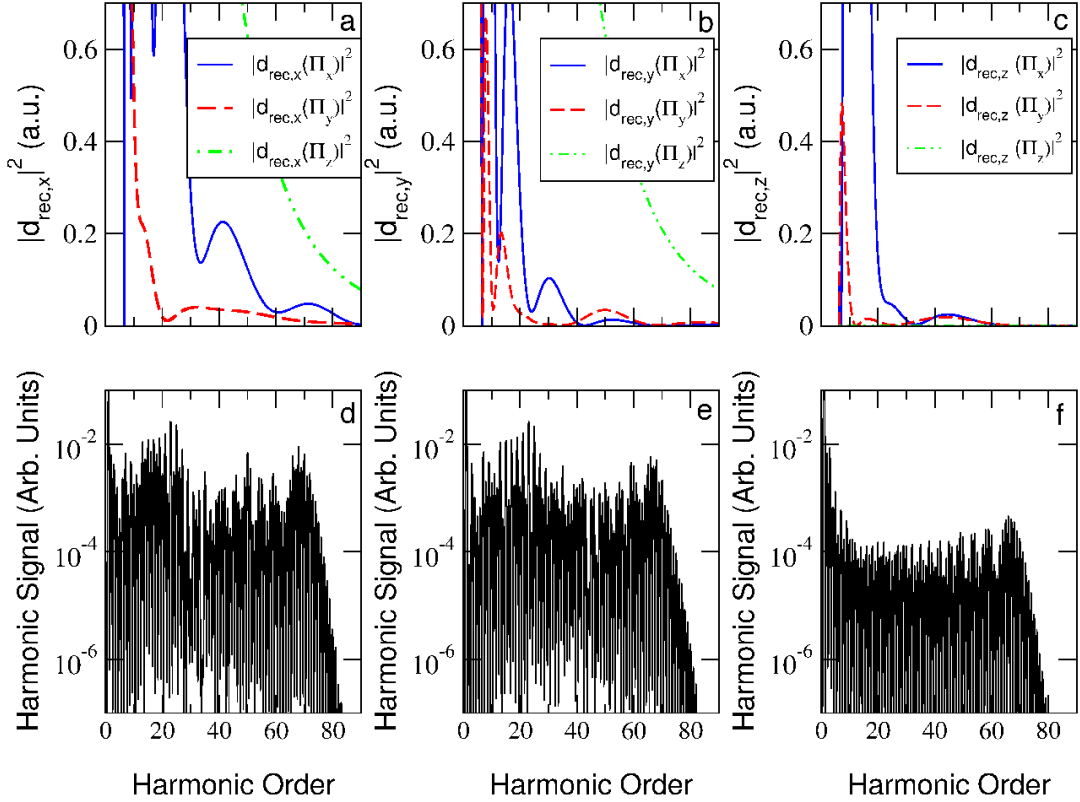}
\caption{(a) Modulus {squared} of the recombination matrix element $x$-component as a function of $\Pi_x$, $\Pi_y$, and $\Pi_z$ calculated for the bowl. HOMO and HOMO-1 contributions have been included. (b) The same as in panel (a) but for the $y$-component of the RME. (c) The same as in panel (a) but for the $z$-component of the RME. (d) Harmonic response $d^{(x)}_{x}$. (e) Harmonic response $d^{(y)}_{y}$. (f) Harmonic response $d^{(z)}_{z}$. }
\label{fig:C20f3_rec_mat_H_H-1}
\end{figure}

 Fig.~\ref{fig:C20f3_rec_mat_H_H-1} shows the modulus {squared} of the three RME components, calculated according to Eq.~\ref{eq43}, as a function of momentum [panels (a), (b), and (c)] and the corresponding harmonic responses 
 [panels (d), (e), and (f)] for the {$C_{20}$} bowl. Comparing those results with the calculations including just the HOMO (Fig.~\ref{fig:C20f3_rec_mat}), 
 it is evident that the RME components as a function of $\Pi_x$ and $\Pi_y$ present a different behavior. This is in contrast to the $\Pi_z$ case which does not show important changes.  The $d^{(z)}_{z}$ harmonic responses [panel (d) in Figs.~\ref{fig:C20f3_rec_mat} and \ref{fig:C20f3_rec_mat_H_H-1}] are almost identical, meaning that in this case  the contribution of HOMO-1 does not play an important role. In the $d^{(x)}_{x}$ and $d^{(y)}_{y}$ cases, the major differences are located between the 10th and 40th harmonic orders,  where the RMEs differ the most. Nevertheless, zeroes and minima of the modulus {squared} of the RME components are generally located at the same positions.

\begin{figure}
\centering
\includegraphics[width=0.8\columnwidth]{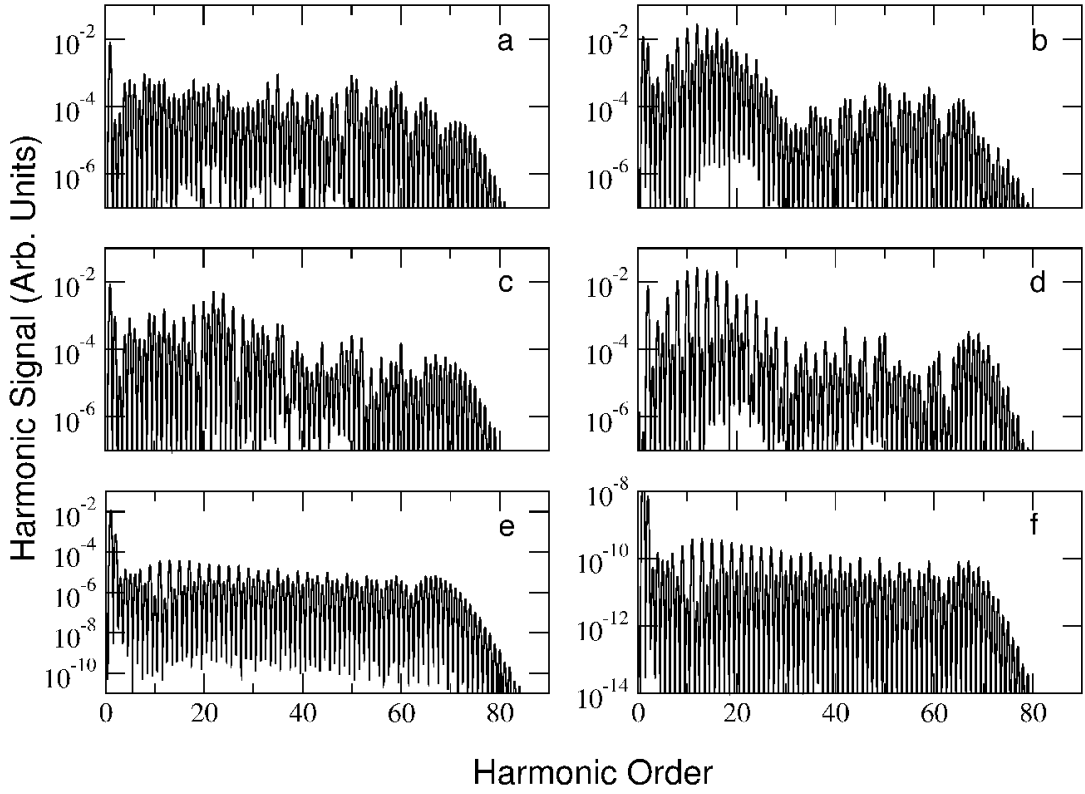}
\caption{HHG spectra calculated for the bowl. Harmonic responses are (a) $d^{(x)}_{y}$, (b) $d^{(x)}_{z}$, (c) $d^{(y)}_{x}$, (d) $d^{(y)}_{z}$, (e) $d^{(z)}_{x}$, and (f) $d^{(z)}_{y}$.}
\label{fig:C20f3_tot_H-H-1}
\end{figure}

Fig.~\ref{fig:C20f3_tot_H-H-1} shows the other harmonic responses obtained from the bowl after the contribution of HOMO-1 is included. Even though the general trend of the plateau in panels (a) through (d) is very similar, some of the peaks present a higher intensity after the last orbital is added. This is particularly true for the region between the 40th harmonic and the cutoff. As expected, the minima seem to be maintained at roughly same locations.  The most important changes are present in panels (e) and (f) in Fig.~\ref{fig:C20f3_tot_H-H-1}. While the responses show a higher intensity, the plateau modulations become less pronounced.

\begin{figure}
\includegraphics[width=0.8\columnwidth]{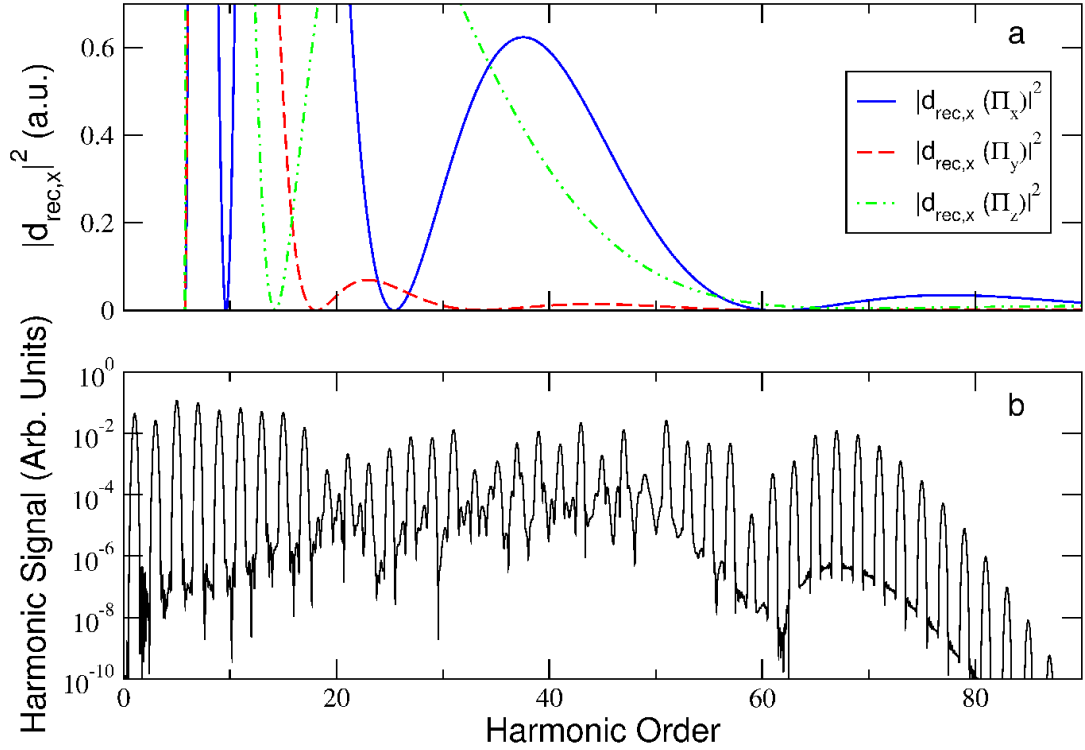}
\caption{(a) Modulus {squared} of the recombination matrix element $x$-component as a function of $\Pi_x$, $\Pi_y$, and $\Pi_z$, calculated for the $C_{20}$ fullerene. The contributions from both HOMO and HOMO-1 have been included. (b) Harmonic response $d^{(x)}_{x}$.}
\label{fig:C20f5_rec_mat_H_H-1}
\end{figure}

Considering now the $C_{20}$ fullerene with the laser field polarized along the $x$-direction, one can clearly see that no important modifications of the RME $x$-component are observed after the addition of the contribution from HOMO-1 (see, Fig.~\ref{fig:C20f5_rec_mat_H_H-1} (a)). When the $\Pi_z$ momentum component is taken into account, a general increase of the RME values is  observed, but the RME zeroes are located at the same positions. The resulting harmonic response $d^{(x)}_{x}$ is almost identical as compared to the HOMO case.  The difference is a small increase in the peak intensity at the beginning of the spectrum for the combined contributions of HOMO and HOMO-1.

\begin{figure}
\includegraphics[width=0.8\columnwidth]{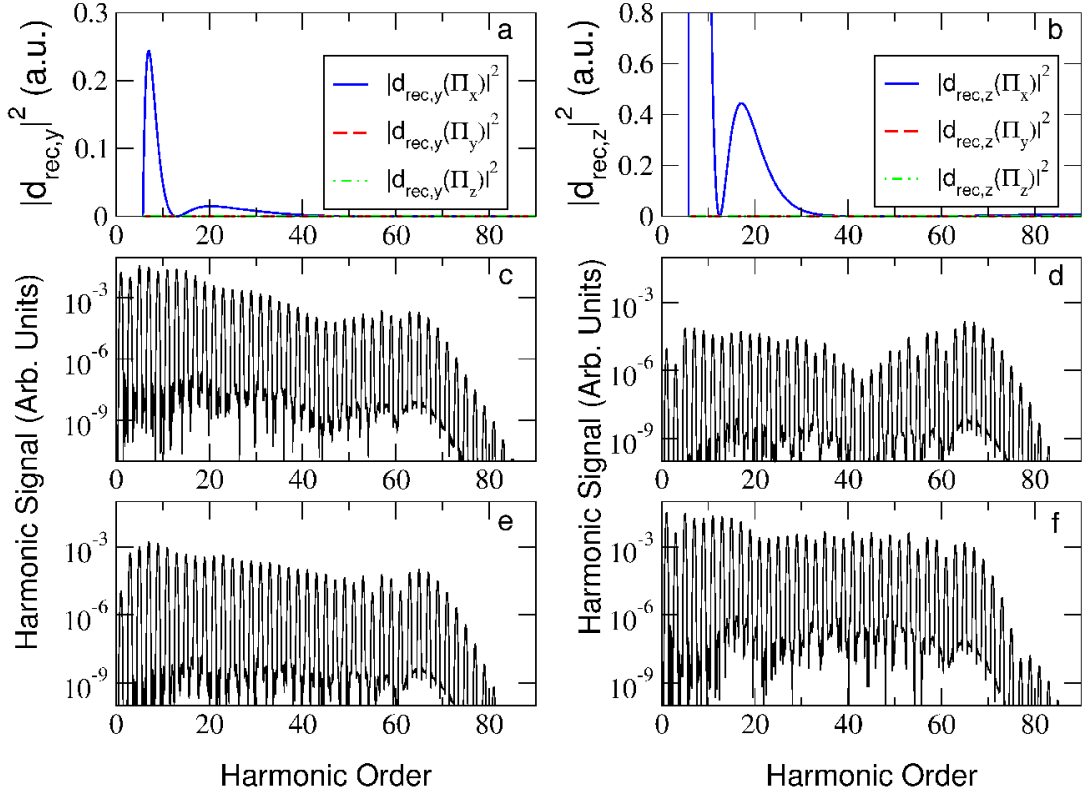}
\caption{Presents the modulus {squared} of the $y$- [panel (a)] and the $z$-component [panel (b)] of the RME and harmonic responses obtained from the $C_{20}$ fullerene, taking into account the contributions from both HOMO and HOMO-1. The laser field is polarized either along the $y$-direction (left column) or along the $z$-direction (right column). The laser field parameters are the same as in Fig.~\ref{fig:C20f5_tot}.  The harmonic responses correspond to $d^{(y)}_{y}$, $d^{(z)}_{y}$, $d^{(y)}_{z}$, and $d^{(z)}_{z}$ (panels (c), (d), (e), and (f), respectively). In contrast to the case when just the HOMO is considered, the contribution from HOMO-1 produces a non-vanishing $y$- and $z$- components of the RME as a function of $\Pi_x$. The remaining RMEs vanish along the spectrum, as in the HOMO case.}
\label{fig:C20f5_tot_H-H-1}
\end{figure}
\begin{figure}
\includegraphics[width=0.75\columnwidth]
{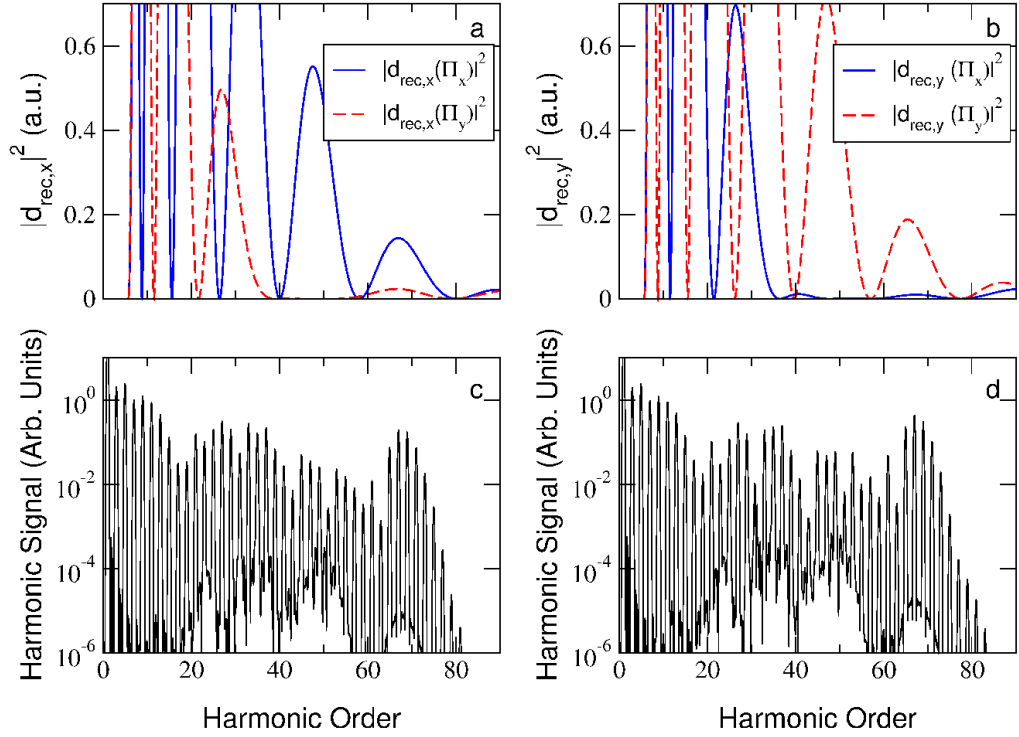}
\caption{Modulus {squared} of the $x$- [panel (a)] and the $y$-component [panel (b)] of the RME as a function of $\Pi_x$ and $\Pi_y$ calculated for the ring including both HOMO and HOMO-1 contributions. The RME as a function of $\Pi_z$ vanishes. (c) Harmonic response $d^{(x)}_{x}$. (d) Harmonic response $d^{(y)}_{y}$. }
\label{fig:C20f2_rec_mat_H_H-1}
\end{figure}

\begin{figure}
\includegraphics[width=0.75\columnwidth]{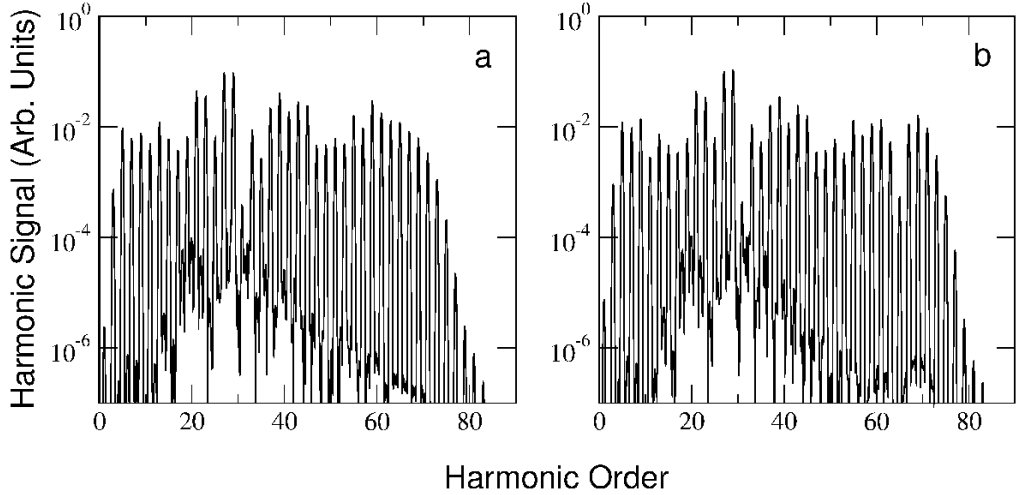}
\caption{HHG spectra from the ring calculated taking into account contributions from both HOMO and HOMO-1. Harmonic responses (a) $d^{(x)}_{y}$ and (b) $d^{(y)}_{x}$.}
\label{fig:C20f2_tot_H-H-1}
\end{figure}
Other harmonic responses from the fullerene are shown in Fig.~\ref{fig:C20f5_tot_H-H-1}. When the HOMO-1 is included in the calculations, the $y$- and $z$- components of the RME as a function of  $\Pi_x$ {acquire} non-zero values (see, panels (a) and (b) of the aforementioned figure). Even though the two RME components which contribute more to the spectral response for the present configurations {[$d_{rec,y}(\Pi_y)$ for a driving field polarized along the $y$-direction and 
$d_{rec,z}(\Pi_z)$ for a driving field polarized along the $z$-direction] vanish, the contribution from ${d_{rec,y}(\Pi_x)}$ or ${d_{rec,z}(\Pi_x)}$} can generate minor changes in the modulations along  the plateau, as it can be seen by comparing Figs.~\ref{fig:C20f5_tot_H-H-1} and \ref{fig:C20f5_tot}. The most important of them is for the $d_{yz}$ harmonic response 
(see panel (b) in Fig.~\ref{fig:C20f5_tot} and panel (d) in Fig.~\ref{fig:C20f5_tot_H-H-1}), where a more pronounced minimum is present at around the 45th harmonic, where the RME $y$- and $z$-components decrease strongly.

\begin{figure}
\includegraphics[width=0.8\columnwidth]{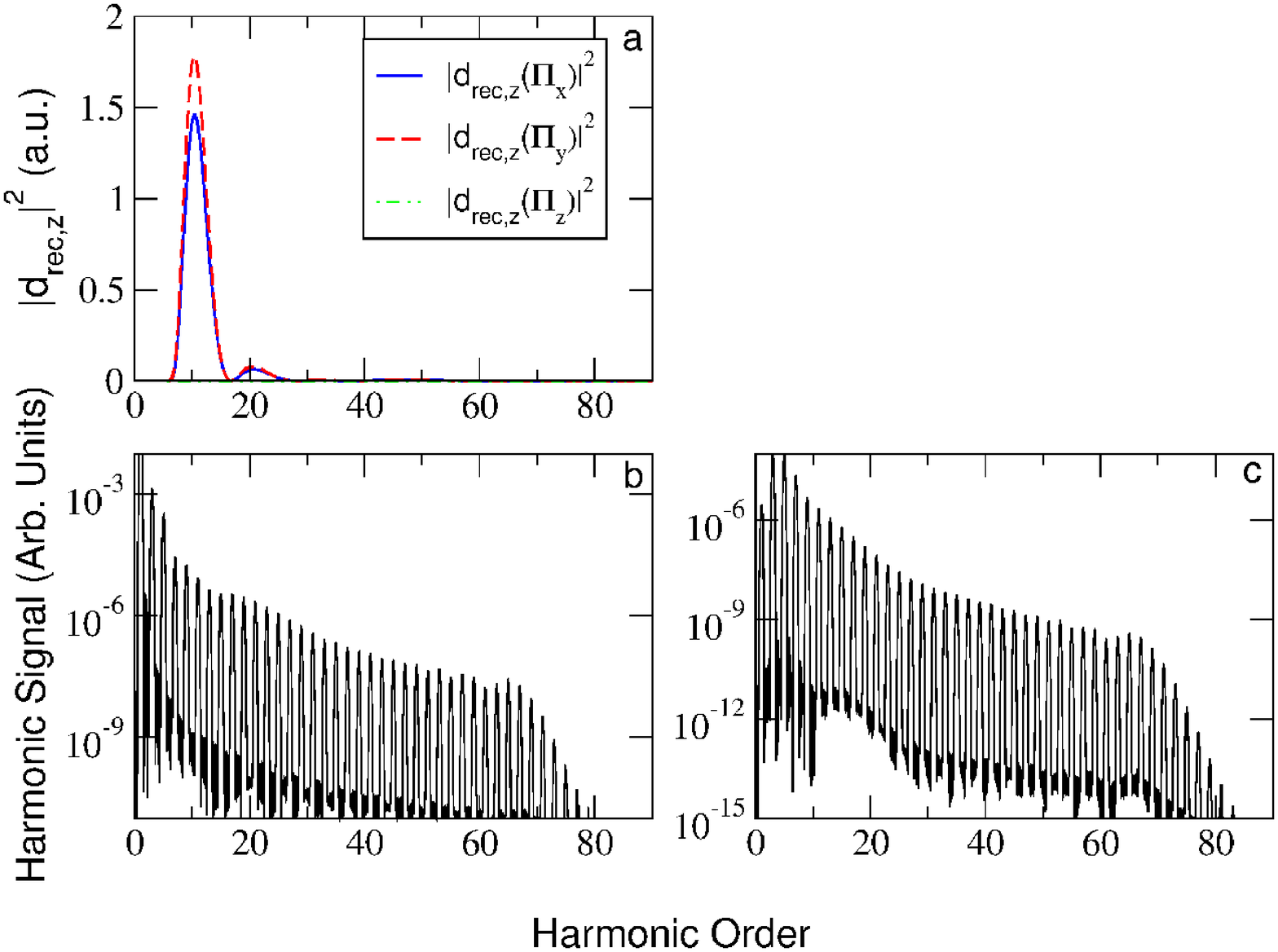}
\caption{Panel (a) presents the modulus squared of the RME $z$-component for the ring when the contributions from HOMO and HOMO-1 are accounted for.  The corresponding spectral response $d^{(z)}_{z}$ is shown in panel (b). In panel (c), we present the harmonic response from the ring restricted to the contribution of the HOMO, for which the corresponding RME component vanishes.}
\label{fig:C20f2_dzz}
\end{figure}

Finally, the inclusion of the HOMO-1 in calculations of the HHG spectra from the ring seems to leave unchanged
the RME $x$- and $y$-components [Figs.~\ref{fig:C20f2_rec_mat_H_H-1} (a) and (b)]. Responses $d^{(x)}_{x}$ and $d^{(y)}_{y}$ [panels (c) and (d), respectively] are identical to the pure HOMO case. The same can be said about the crossed terms $d^{(x)}_{y}$ and $d^{(y)}_{x}$ (see, Fig.~\ref {fig:C20f2_tot_H-H-1}).

When the $d^{(z)}_{z}$ harmonic response from the ring includes the contribution of both HOMO and HOMO-1 orbitals, the $z$-component of the RME as a function of $\Pi_x$ and $\Pi_y$ acquire nonvanishing values, contrary to the case when just the HOMO is considered (see, Fig.~\ref{fig:C20f2_dzz}). 
For the latter, the $z$-component of the RME vanishes for $\Pi_x$, $\Pi_y$, and $\Pi_z$. Even though, the RME which contributes the most to the present configuration [$d_{rec,z}(\Pi_z)$] is zero, it is expected that the two 
other components change in a minor way the shape of the plateau. As it can be seen in Fig.~\ref{fig:C20f2_dzz}, the harmonic response obtained by accounting for HOMO and HOMO-1 [panel (b)] presents a very similar trend to the spectral response obtained from HOMO [panel (c)] with a general increase in intensity. The differences of the modulations between those two cases are minor and appear in the lower energy part of the plateau, when the $z$-component of the RME acquires non-zero values.

\section{Conclusions}

The Lewenstein model is a very useful tool to analyze the harmonic response from atoms and molecules interacting with strong laser fields. We corroborate that the length gauge formalism applied to molecules with small $Q$ values does not involve an unphysical extension of the plateau, as it was pointed out in other works~\cite{Chirila,Ciappina_JB,Ciappina_JB2}.

Our present calculations, in the length gauge, predict harmonic responses with different polarization directions, which depend on the particular isomer and driving field polarization.
We have shown that different molecular arrangements, as in the case of the three $C_{20}$ isomers, lead to different spectral responses. Multi-slit interference patterns, which produce intensive modulations of the harmonic responses along the plateau, are related to the nuclear distribution in the molecule and its molecular orbital configuration. The zeroes of the recombination matrix elements as a function of momentum in the three coordinates are closely related to the interference effects evidenced as minima in the plateaus.

We have shown that some of the harmonic responses from the $C_{20}$ fullerene, for which the RME is not oscillating, present modulations in peak intensity along the plateau. Such modulations are related to quantum path interferences, which happen when  electron wave packets following different trajectories interfere.

We believe that the observation of the harmonic polarization direction, together with the analysis of multi-center interference minima can help in the differentiation between different aligned harmonic targets. Those properties can be used in the development of a simple spectroscopic technique.

\section{Acknowledgements}

A. J. acknowledges support from U.S. National Science Foundation (Grant No. PHY-1734006 and Grant No. PHY-2110628).

\end{document}